\documentclass[12pt,preprint]{aastex}

\slugcomment{Accepted to ApJ}

\shorttitle{Survey of Embedded Clusters in $\CO$($J$=1--0) line emission}
\shortauthors{A.E. Higuchi et al.}

\def\MO{M_\odot}
\def\CO{\mathrm{C}^{18}\mathrm{O}}
\def\CCO{\mathrm{C}^{17}\mathrm{O}}
\def\CCCO{^{13}{\rm{C}}\rm{O}}
\def\MVIR{M_{\mathrm{VIR}}}
\def\MCLU{M_{\mathrm{clump}}}
\def\dv{\Delta{V}}
\def\dvC{\Delta{V}_{\mathrm{clump}}}
\def\RC{R_{\mathrm{clump}}}

\begin{document}

\title{A Mapping Survey of Dense Clumps Associated with Embedded Clusters
: Evolutionary Stages of Cluster-Forming Clumps}

\author{Aya E. HIGUCHI\altaffilmark{1,2}, Yasutaka KURONO\altaffilmark{3}, Masao SAITO\altaffilmark{3}, 
\and  Ryohei KAWABE\altaffilmark{2}}
\email{higuchi@nro.nao.ac.jp}

\altaffiltext{1}{Department of Earth and Planetary Sciences Tokyo Institute of Technology 2-12-1 Ookayama, Meguro-ku, Tokyo 152-8551, Japan}
\altaffiltext{2}{Nobeyama Radio Observatory, Nobeyama, Minamimaki, Minamisaku, Nagano 384-1305, Japan}
\altaffiltext{3}{National Astronomical Observatory of Japan 2-21-1 Osawa, Mitaka, Tokyo, 181-8588, Japan}

\begin{abstract}

We have carried out a survey of the dense clumps associated with 14 embedded clusters in the $\CO$ ($J$=1--0) line emission with the Nobeyama 45m telescope in order to understand the formation and evolution of stellar clusters in dense clumps of molecular clouds.
We have selected these clusters at distances from 0.3 to 2.1 kpc and have mapped about 6$^{\prime} \times 6^{\prime}$ to 10$^{\prime} \times 10^{\prime}$ regions (corresponding to 3.8$\,$pc $\times$ 3.8$\,$pc at 2.1$\,$kpc) for all the clumps with 22$''$ resolution (corresponding to Jeans length at 2.1$\,$kpc). 
We have obtained dense clumps with radii of 0.40--1.6$\,$pc, masses of 150--4600$\,$$\MO$, and velocity widths in FWHM of 1.4--3.3$\,$${\rm{km \: s}^{-1}}$.
Most of the clumps are found to be approximately in virial equilibrium, which implies that $\CO$ gas represents parental dense clumps for cluster formation.
From the spatial relation between the distributions of clumps and clusters, we classified $\CO$ clumps into three types ($\it{Type \ A, \ B, \rm{and}}$ $\it{C}$).
The $\CO$ clumps as classified into $\it{Type \ A}$ have emission distributions with a single peak at the stellar clusters and higher brightness contrast than that of other target sources. 
$\it{Type \ B}$ clumps have double or triple peaks which are associated with the cluster and moderately high brightness contrast structure. $\it{Type \ C}$ clumps have also multiple peaks although they are not associated with the cluster and low brightness contrast structure.
We suggest that our classification represents an evolutionary trend of cluster-forming dense clumps because dense gas in molecular clouds is expected to be converted into stellar constituents, or to be dispersed by stellar activities. 
Moreover, although there is a scatter, we found a tendency that the SFEs of the dense clumps increase from $\it{Type \ A}$ to $\it{Type \ C}$, which also supports our scenario.
\end{abstract}


\section{INTRODUCTION}

Star formation is one of the most important issues to be understood in astronomy.
It has been accepted that the star formation processes are roughly classified into two modes (e.g., $\,$ Carpenter 2000). 
One is the isolated star formation in which mode low-mass ($<$$\,$2$\,$$\MO$) stars form inside isolated dense ($\sim$10$^5$$\,$cm$^{-3}$) cores with sizes of $\sim$$\,$0.1 pc as typically seen in the Taurus molecular cloud \citep{oni02, tat04}.
The other mode corresponds to the cluster formation in which low- to high-mass stars form as a cluster, as in the Orion molecular cloud. 
Most stars (more than 90$\,$$\%$ of stars within our galaxy), particularly massive stars ($>$$\,$8$\,$$\MO$), within the disk of the Milky Way form as member of clusters \citep{lad03}. Thus, clusters have been long considered as important laboratories for astronomy.
Recent studies showed that clusters form within dense clumps (size : $\sim$1 pc, mass : 100--1000$\,$$\MO$, density : 10$^{3-5}$$\,$$\rm{cm}^{-3}$) which are considered to be the parental objects of clusters \citep{lad03}.
However, there are still a small number of systematic studies about the relations between the dense clumps and clusters.
In order to understand the physical processes of the clusters, one of the important approaches is to investigate the physical conditions of clumps and clusters along the evolutionary stages. 
Thus, a comprehensive survey of cluster-forming regions, to reveal the physical properties of dense gas and stellar association, plays crucial roles to understand the cluster formation processes including the initial conditions.

In previous studies, Ridge et al. (2003) surveyed 30 cluster-forming clumps using $^{13}\rm{CO}$($J$=1--0), $\CO$($J$=1--0), and $\CO$($J$=2--1) molecular lines.
They classified the clumps into three stages, $\it{Class \ I, \ II, \rm{and}}$ $\it{III}$ from
the morphologies, peak numbers of the clumps, and the position of $\it{IRAS}$ or near-infrared (embedded) sources.
$\it{Class \ I}$ clumps have a single peak in $\CO$ and $\CCCO$ which is associated with a cluster, and by contrast
$\it{Class \ III}$ clumps have multiple peaks in $\CCCO$ and multiple peaks or no emission in $\CO$ which are not associated with a cluster.
The rest clumps are classified into $\it{Class \ II}$ which have filamentary structures and multiple peaks in $\CO$ and $\CCCO$.
In $\it{Class \ II}$ clumps, the cluster is associated with one of multiple peaks.
They proposed an evolutionary scenario of the dense gas from $\it{Class \ I}$ to $\it{Class \ III}$ clumps.
However, there are still remaining problems.
At first, the spatial resolution is not enough to investigate gas dispersal and to compare the spatial relation between the observed clump distributions and the embedded stellar clusters.
They referred to stellar information only from the $\it{IRAS}$ data with worse resolution than that of 2MASS or $\it{Spitzer}$ data.
Second, their target sources contain evolved clusters that it is difficult to examine physical relations between the dense gas and formed clusters. On the other hand, Fuente et al. (1998a, 2002) revealed the progressive dispersal of the dense gas associated with Herbig Ae/Be stars in their evolution to the main sequence (e.g., Saito et al. 2001 for low mass stars). 
From the observations using $^{13}\rm{CO}$($J$=1--0), and CS($J$=3--2) molecular lines, and 1.3 mm dust continuum,
they also proposed an evolutionary classification of the parental gas surrounding a central star from morphologies.
However, they only mapped $\sim$ 1$\,$pc$^{2}$ regions around the Herbig Ae/Be stars.
From the current situation, we considered that observational studies toward on-going cluster-forming clouds using molecular lines sensitive to dense cloud gas, using high-resolution and wide-field imaging, are required.

In this paper, we present high-resolution $\CO$($J=$1--0) molecular line data observed with the Nobeyama 45m telescope toward 14 clusters within 2.1$\,$kpc.
Six among our targets are overlapped with samples by Ridge et al. (2003) who studied only nearby targets located within 1$\,$kpc.
We morphologically classify the cluster-forming clumps according to the spatial relation between the distributions of clumps and clusters, and discuss the evolution of dense clumps in course of cluster formation.

\section{OBSERVATIONS}

\subsection{The Nobeyama 45m Telescope Observations}

We have carried out $\CO$($J$=1--0; 109.782182 GHz) observations using the 45m telescope in the Nobeyama Radio Observatory (NRO) from 2005 December to 2008 May. 
At 110 GHz the telescope has a beamsize in full width at half maximum (FWHM) of 15$^{\prime\prime}$, corresponding to the spatial resolution of 0.2 pc at a distance of 2.1$\,$kpc, and the main beam efficiency, $\eta$ of 0.45.
We used the 25-BEam Array Receiver System (BEARS) at the front end, which has 5 $\times$ 5 beams separated by 41$^{\prime\prime}$.1 on the plane of the sky \citep{sun00, yam00}. At the back end, we used 25 sets of 1024 channel Auto-Correlators (ACs), which have 32 MHz bandwidth and 37.8 kHz frequency resolution \citep{sor00}. The frequency resolution corresponds to the velocity resolution of 0.10$\,$km $\rm{s}^{-1}$ at 110 GHz.
During the observations, the system noise temperatures were in the range between 250 to 450$\,$K in double side band. 
We used the emission free area near the observed sources as the off positions. 
The standard chopper wheel method was used to convert the calibrated intensity into the antenna temperature 
${T}^{*}_{\mathrm{A}}$, corrected for the atmospheric attenuation. 
The telescope pointing was checked every 1.5 hours by observing SiO maser sources near the target objects. 
The pointing accuracy was better than 3$^{\prime\prime}$.
To correct the difference of intensity scales among the 25 beams of the BEARS, we used calibration data obtained from the observations
toward the center of S140 region using another SIS receiver, S100 with a SSB filter and Acousto-Optical Spectrometers (AOSs) as a backend.
Our mapping observations were made in the On-The-Fly (OTF) mapping technique \citep{saw08} and covered approximately 6$^{\prime}$ $\times$ 6$^{\prime}$ to 10$^{\prime} \times 10^{\prime}$ areas (corresponding to 3.8$\,$pc $\times$ 3.8$\,$pc at a distance of 2.1$\,$kpc) centered on the position of the clusters. 
We combined scans along R.A. and decl., to reduce the scanning effects.

We used a convolutional scheme with a Spheroidal function \citep{saw08} to calculate the intensity at each grid point of the final 
map-cube ($\alpha$, $\delta$, and $v_{\mathrm{LSR}}$) data with a spatial grid size of 7$^{\prime\prime}$.5, half of the beamsize: the final effective resolution became 22$^{\prime\prime}$ and the effective integration time at each grid point was about 80 sec. 
As a result, a mean rms noise level of all the objects became 0.11 K in ${T}^{*}_{\mathrm{A}}$.
To estimate the optical depth of the $\CO$($J$=1--0) emission, one-point $\CCO$($J$=1--0; 112.358988 GHz) observations were carried out toward a peak of the $\CO$ emission in Mon R2, which has the strongest $\CO$ emission in our targets, with the 45m telescope.
The beamsize in FWHM and $\eta$ were 15$^{\prime\prime}$ and 0.45, respectively.
We used the S100 receiver at the front end. 
As the back end, we used the AOSs, which have the velocity resolutions of 0.10$\,$km s$^{-1}$ at 112$\,$GHz.
The system noise temperatures were around 450$\,$K. 
The integration time was 30 minutes, resulting in the rms noise level of 0.05 K in ${T}^{*}_{\mathrm{A}}$. 
The telescope pointing was checked every 1.5 hours by observing the SiO maser source in the Orion KL. 
The pointing accuracy was better than 3$^{\prime\prime}$.

\subsection{Source selection}

We selected 14 nearby embedded clusters from the catalogue by Lada $\&$ Lada (2003) and Porass et al. (2003).
The criteria for our target selection are to be within 2.1$\,$kpc from the Sun, to be associated with natal clumps.
The most massive stellar mass of each of target source has been well determined.
In addition, the target sources are found to be associated with NH$_{3}$ emission, suggestive of being relatively young.
The NH$_{3}$ detection indicates the presence of dense gas, 
i.e., the natal clump, and strongly suggests that they are on-going cluster formation regions.
We exclude the Orion nebula cluster which has been well studied (e.g.,$\,$Ikeda et al. 2007) and objects 
in the southern hemisphere which cannot be observed in Nobeyama.

\begin{figure}
\epsscale{0.7}
\plotone{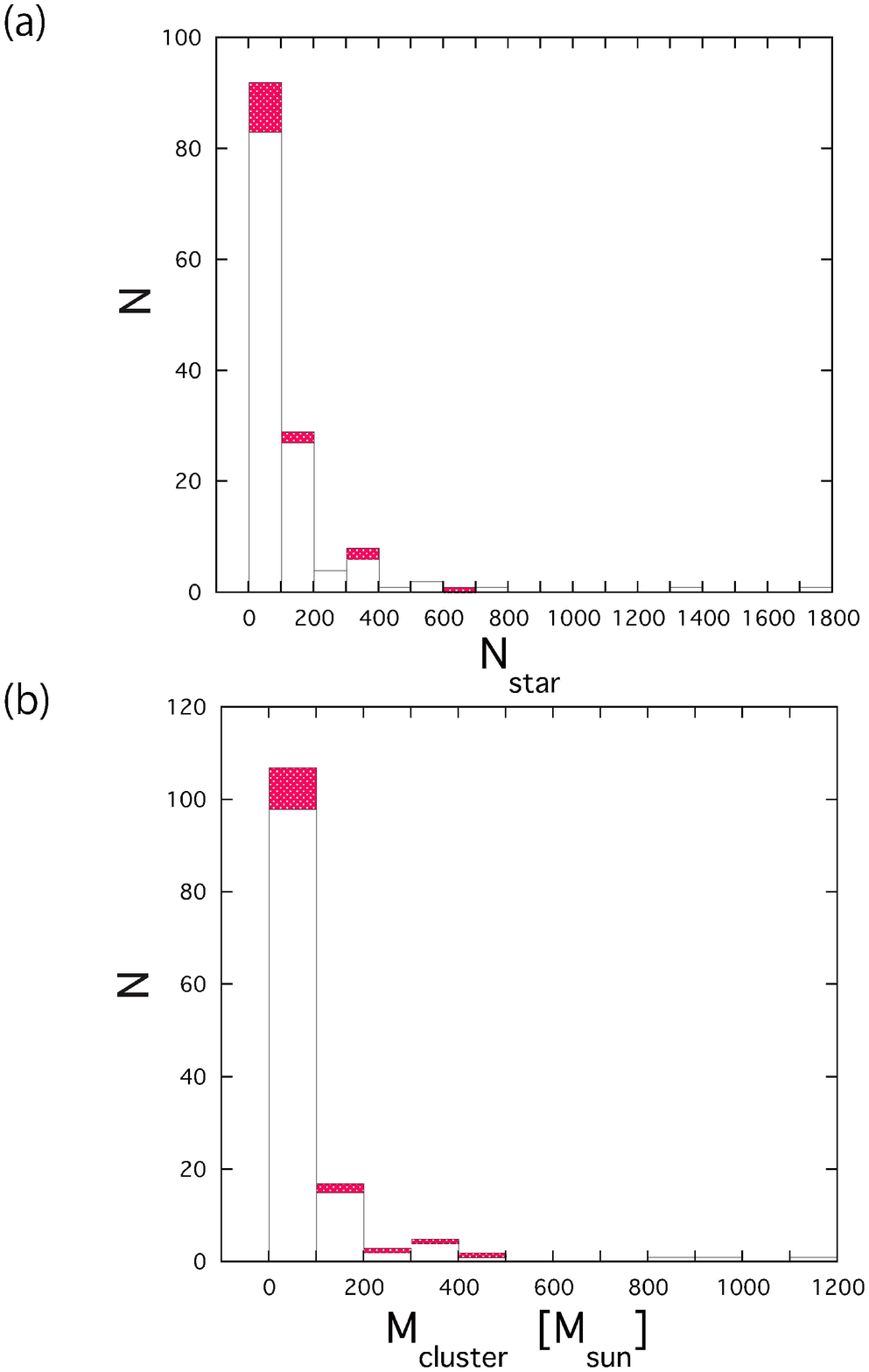}
\caption{Histograms of stellar number $N_{\mathrm{star}}$ (a), 
and mass $M_{\mathrm{cluster}}$ (b) of clusters in catalog by Lada $\&$ Lada (2003). 
Our target is hatched area. The detailed definition is written in Lada $\&$ Lada (2003).}
\label{fig1}
\end{figure}

The physical properties of our target clusters are summarized in Table \ref{target}, the total stellar masses of the clusters (12--340$\,$$\MO$), the number of the member stars (16--550), and the highest stellar mass of the members (3--20$\,$$\MO$), most of which are derived from the near infrared observations (e.g., 2MASS). Only for S140 and BD+40$^{\circ}$4124, the total masses of the stellar members are estimated using the $N_{\rm{star}}$-$M_{\rm{cluster}}$ relation that was derived using the data in Lada $\&$ Lada (2003).
Thirteen among our targets are massive star forming regions, whose maximum stellar masses are greater than or equal to $\sim$ 8$\,$$\MO$, and half of the massive star forming regions have H {\sc ii} regions.
Figure \ref{fig1} shows the statistics of stellar number, and mass of the selected 14 clusters (hatched) and the other objects listed in Lada $\&$ Lada (2003) (white).
Our selected objects are expected to be young cluster-forming regions unbiased in physical scales, and suitable for our study of cluster formation.
We used NH$_{3}$($J,K$=1,1) and (2,2) data taken with Nobeyama 45m telescope from the NRO archive system.
The spatial resolution was 73$''$, and the typical noise level is 0.05$\,\rm K$ with a velocity resolution of 0.5$\,$km s$^{-1}$.

\section{RESULTS}

\subsection{Total integrated intensity and velocity dispersion maps in $\CO$}\label{3-1}

\begin{figure}
\epsscale{0.8}
\plotone{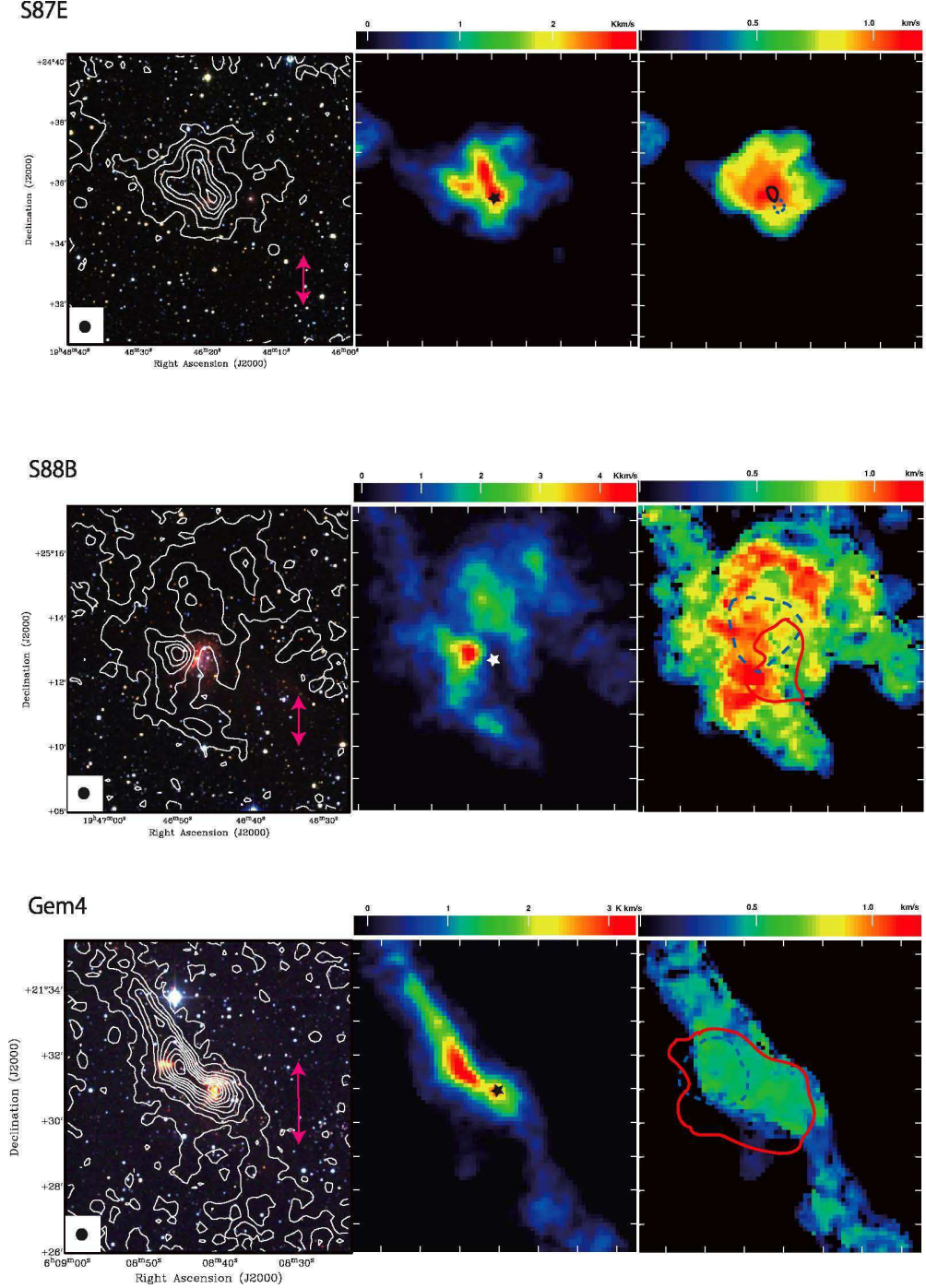}
\caption{
Total integrated intensity maps of the $\CO$($J$=1--0) emission 
(contours) superposed on the $JHK_{\rm{s}}$ composite color images in log scale from 2MASS (left),
total integrated intensity maps (middle), and velocity dispersion maps (right) for
S87E (top), S88B (middle), and Gem 4 (bottom). 
The contours with the intervals of the 5 $\sigma$ levels start from 
the 5 $\sigma$ levels, where the 1 $\sigma$ noise levels are 0.09$\,$K km s$^{-1}$, 0.16$\,$K km s$^{-1}$, and 0.07$\,$K km s$^{-1}$
in ${T}^{*}_{\mathrm{A}}$ for S87E, S88B, and Gem 4, respectively.
The filled circle at the bottom left corner in each panel shows the effective resolution in FWHM of $22^{\prime\prime}$.
The outlines of the blue- and red-shifted outflows are also plotted on the velocity dispersion maps.
The stellar marks indicate the position of cluster center. 
The pink arrows show 1 pc scale of individual regions.
The 2nd moment maps were spatially smoothed to enhance S/N ratio with an effective resolution of approximately 40$''$.}
\label{map1}
\end{figure}

\begin{figure}
\epsscale{0.9}
\plotone{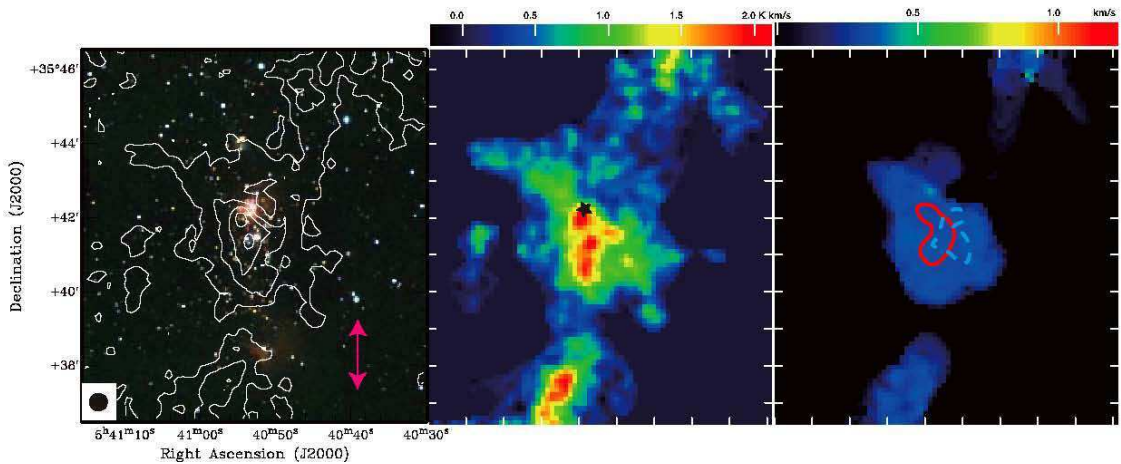}
\caption{
Same as Figure \ref{map1} for AFGL 5180 (top), GGD12-15 (middle), and S235AB (bottom).
The 1 $\sigma$ noise levels are 0.07$\,$K km s$^{-1}$, 0.12$\,$K km s$^{-1}$, and 0.10$\,$K km s$^{-1}$ in ${T}^{*}_{\mathrm{A}}$ for AFGL 5180, GGD12-15, and S235AB respectively.}
\label{map2}
\end{figure}

\begin{figure}
\epsscale{0.9}
\plotone{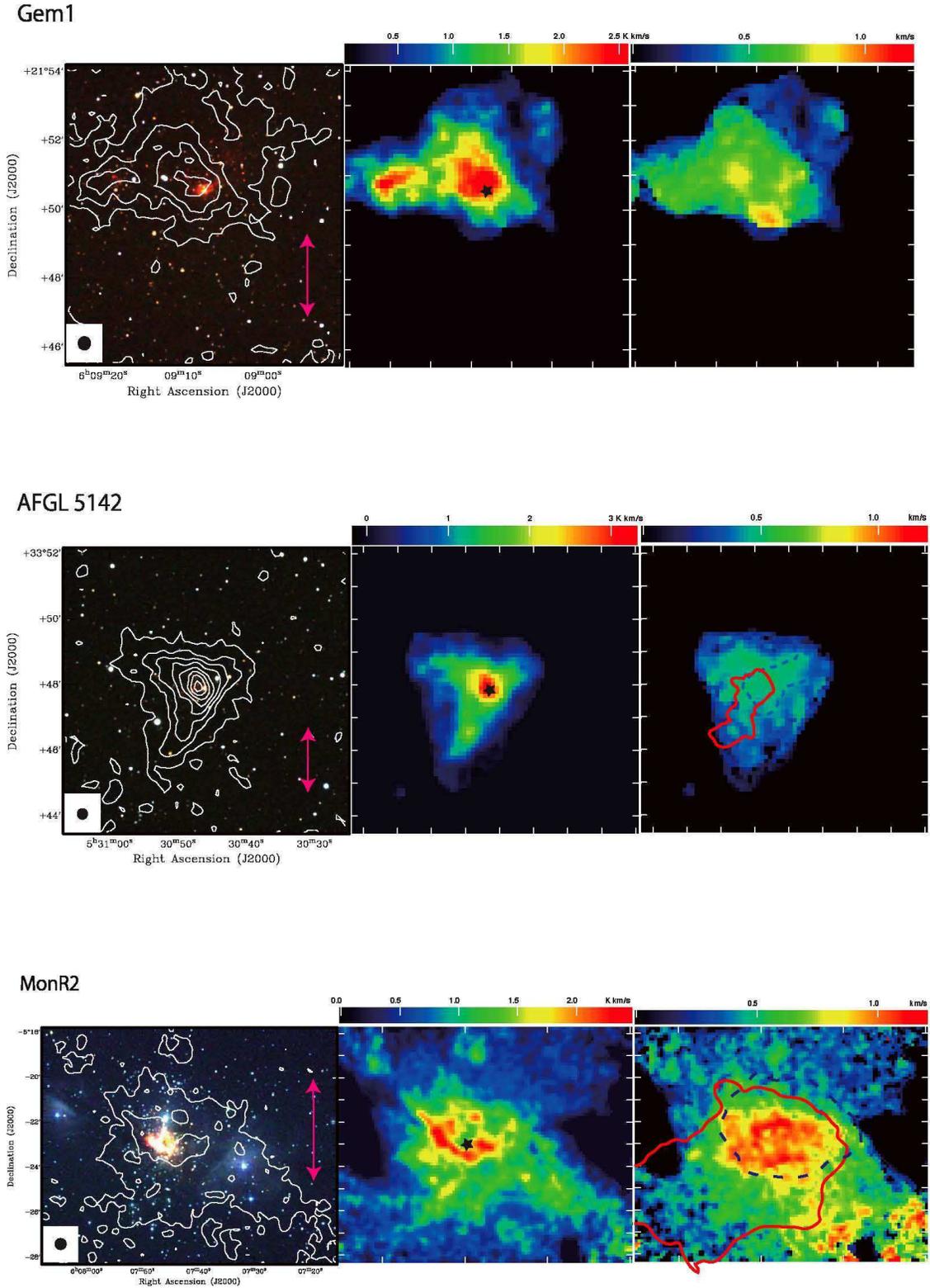}
\caption{Same as Figure \ref{map1} for Gem 1 (top), AFGL 5142 (middle), and Mon R2 (bottom).
The 1 $\sigma$ noise levels are 0.12$\,$K km s$^{-1}$, 0.08$\,$K km s$^{-1}$, and 0.15$\,$K km s$^{-1}$ in ${T}^{*}_{\mathrm{A}}$ for Gem 1, AFGL 5142, and Mon R2, respectively.
\label{map3}}
\end{figure}

\begin{figure}
\epsscale{0.9}
\plotone{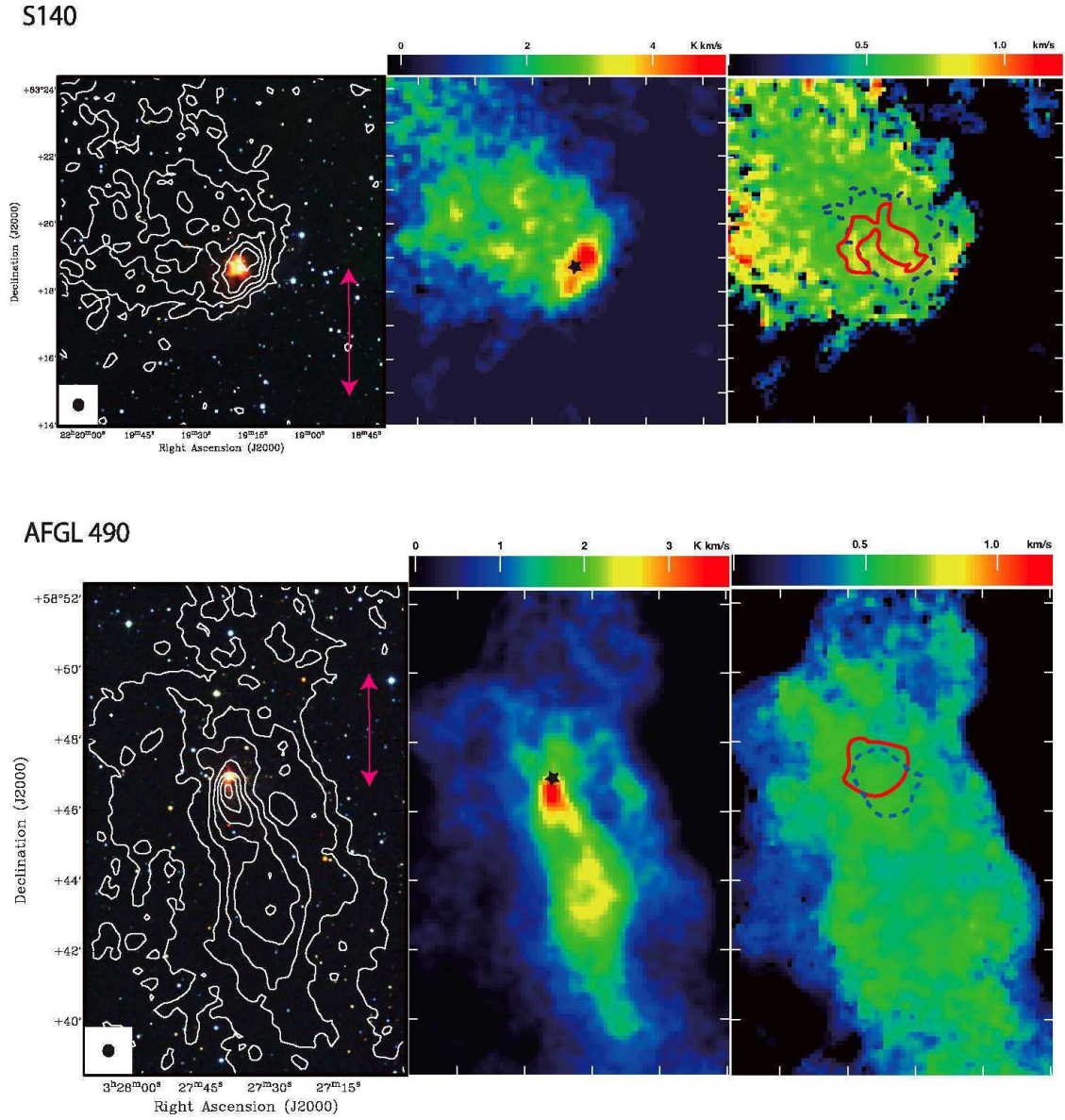}
\caption{Same as Figure \ref{map1} for S140 (top) and AFGL 490 (bottom).
The 1 $\sigma$ noise levels are 0.18$\,$K km s$^{-1}$, and 0.13$\,$K km s$^{-1}$ in ${T}^{*}_{\mathrm{A}}$ for S140 and AFGL 490, respectively.
\label{map4}}
\end{figure}

\begin{figure}
\epsscale{0.9}
\plotone{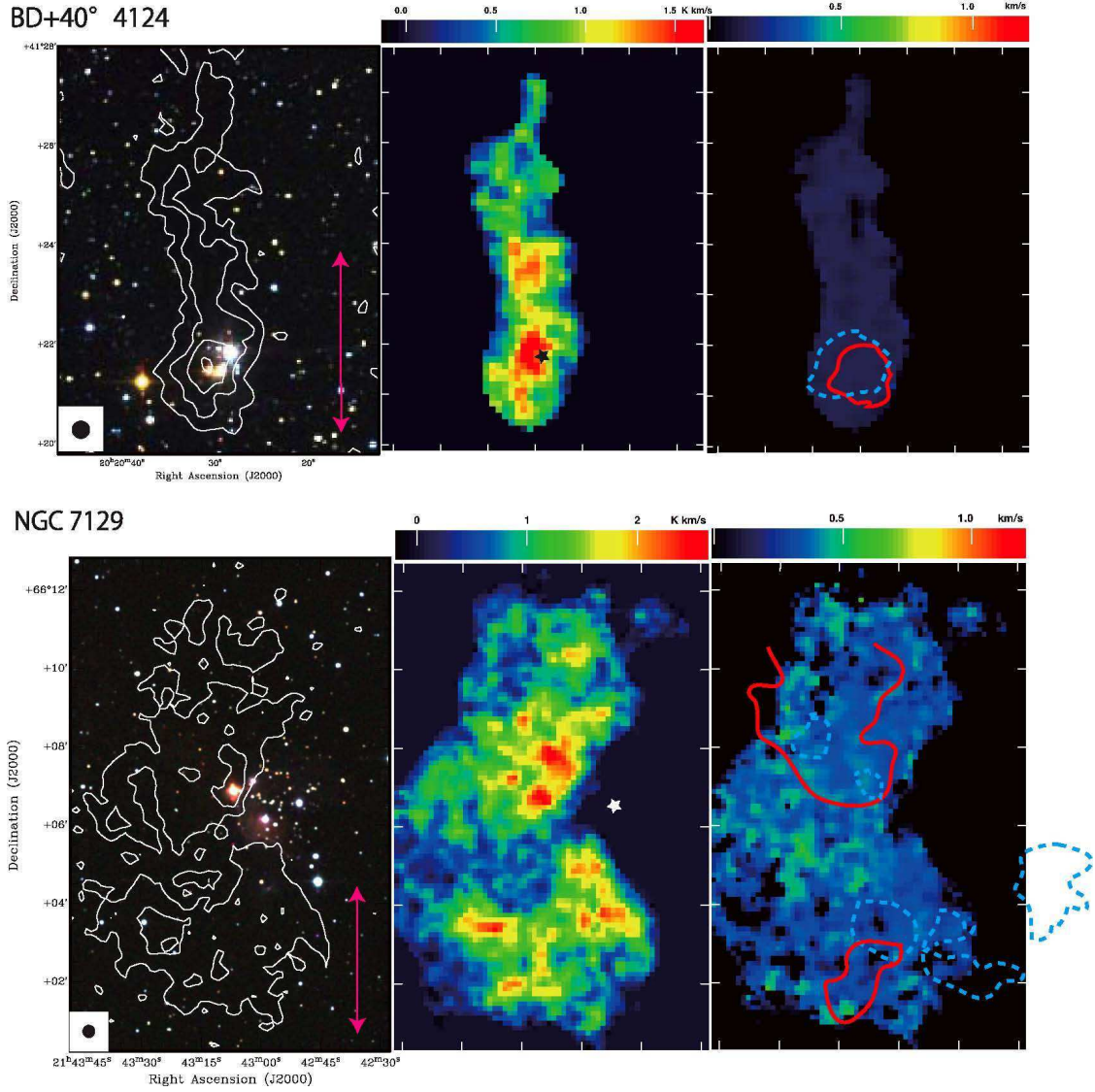}
\caption{Same as Figure \ref{map1} for BD+40$^{\circ}$4124 (top), NGC 7129 (bottom).
The 1 $\sigma$ noise levels are 0.08$\,$K km s$^{-1}$, and 0.16$\,$K km s$^{-1}$ in ${T}^{*}_{\mathrm{A}}$ for BD$+$40$^{\circ}$4124, NGC 7129, respectively.  
\label{map5}}
\end{figure}

\begin{figure}
\epsscale{0.9}
\plotone{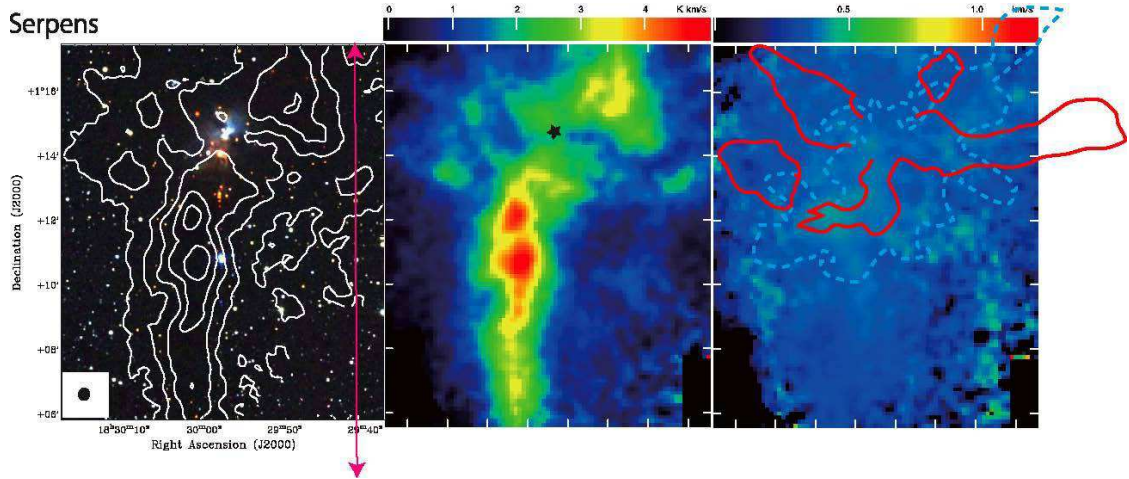}
\caption{Same as Figure \ref{map1} for Serpens (bottom).
The 1 $\sigma$ noise levels are 0.13$\,$K km s$^{-1}$ in ${T}^{*}_{\mathrm{A}}$ for Serpens.  
\label{map6}}
\end{figure}

The left panels of Figures \ref{map1} to \ref{map6} show the $\CO$ total integrated intensity maps with about 6$^{\prime} \times 6^{\prime}$ to 10$^{\prime} \times 10^{\prime}$ size superposed on the 2MASS images ($J,H, \ $\rm{and}$ \ {K}_{\rm{s}}$ composite color images).
The detailed derivations of the physical properties are described in $\S$ \ref{3-2}.
The $\CO$ clumps show various morphologies, compact or extended, spherical, cometary, or filamentary.
Moreover, our high-resolution maps show that most of the clumps have internal structures which consist of several emission components.
We defined the internal components whose separations are comparable to the order of Jeans length as the sub-clumps. 
The Jeans length, $\lambda_{\rm{J}}$ was calculated as
$\lambda_{\rm J}=0.27\,{\rm pc}\left(T_{\rm K}/20\,{\rm K}\right)^{1/2}\left(n_{\rm H_2}/10^4\, {\rm cm^{-3}}\right)^{-1/2}$
, where $T_{\rm{K}}$ is the kinetic temperature of the clump and $n_{\rm H_{2}}$ is the density of the clump.
We derived the kinetic temperature, $T_{\rm{K}}$ in $\S$ \ref{3-2}.
The density of the clump, $n_{\rm H_{2}}$ was calculated as $n_{\rm H_{2}}=({3/4\pi})({\MCLU}/\mu{\rm{m_{H_{2}}}}{\RC}^{3})$, 
where $\MCLU$ is LTE mass of the clump, $\mu{\rm{m_{H_{2}}}}$ is mean mass of molecule, and $\RC$ is the radius of the clump derived in $\S$ \ref{3-2}.

For each target source, we defined the cluster center, $\CO$ emission peak position (multiple peaks in some cases), and $\CO$ cavity for comparison as follows. We defined the cluster center as the position of the most massive star, the peak positions as positions where $\CO$ column density is locally enhanced by a factor of a few compared to the mean value of the clump.
For the cavity, we also defined as a place where the column density is decreased by more than a factor of three. 
We categorize the $\CO$ peak positions into two cases, ``Peak" and ``Off-peak", using the separation between the $\CO$ peak positions and the cluster centers; if the separation is smaller than the Jean length estimated for each target we call ``Peak" otherwise ``Off-peak".
In addition, we can see a variation in the relative positions of stellar clusters in the distributions of $\CO$ emission. 
For example, while for AFGL 5142 the cluster members are just located at the single peak position of the $\CO$ 
emission, for S88B the stellar components are not located at the emission peaks but at emission valleys.
The middle panels of Figures \ref{map1} to \ref{map6} also show the total integrated intensity maps.
The right panels of Figures \ref{map1} to \ref{map6} show the $\CO$ velocity dispersion maps, namely intensity weighted 2nd moment maps of $\CO$ produced 
in AIPS\footnote{The National Radio Astronomy Observatory is a facility of the National Science Foundation operated under cooperative agreement by Associated Universities, Inc.}. 
The velocity dispersion is defined as $\int{I(v-v_{0})^{2}dv}/\int{Idv}$, where $\int{Idv}$ is the total integrated intensity of the map,
$v_{0}$ is the peak velocity of the pixel in the map.
We note that the 2nd moment maps were smoothed to enhance a signal to noise ratio. 
As a result, the effective resolution of these maps are approximately 40$''$.
Here, we used the 2nd moment map to compare the spatial distribution of the velocity dispersion and outflows, the locations of clusters in $\S$ \ref{4-2-3}.
It is found that most of the $\CO$ clumps have relatively flat distributions in
velocity dispersion within the clumps except for S87E, S88B, and Mon R2.
These three objects have regions of large velocity dispersion toward the cluster centers, 
up to $\sim$ 1.3$\,$km s$^{-1}$ (averaged value for the clumps which we obtained is 0.7$\,$km s$^{-1}$).
Most of our target cluster-forming regions were found to have molecular outflows in previous studies (e.g.,$\,$Wolf et 
al. 1990 for Mon R2) as shown in the right panels of Figures \ref{map1} to \ref{map6}, which is important information for inclusive consideration and understanding of cluster formation. 
The detailed clump features of the individual regions are presented in Appendix A.

\subsection{Physical properties of the $\CO$ clumps}\label{3-2}

In this section we describe the derivation of the physical properties of the $\CO$ clumps.
We summarize the physical properties in the following methods in Table \ref{para}.
Hereafter, we call dense clumps traced by the $\CO$ line emission as $\CO$ clumps whose intensities are above the 5 $\sigma$ noise levels in the individual maps.

To define the clump radius, we estimated the projected area of the clump on the sky plane, $A$ and the observed radius
was derived as $R_{\rm{obs}}$=$(A/\pi)^{1/2}$.
Then we corrected for the spatial resolution as, 
\begin{eqnarray}
\label{radius}
R_{\rm{clump}} \quad=\quad [ R_{\rm{obs}}^2 -
		(\theta_{\rm FWHM}/2)^2]^{1/2} ,
\end{eqnarray}
where $\theta_{\rm{FWHM}}$ of $22''$ the effective beam size in FWHM.

The velocity width of the clump (one-dimensional velocity widths in FWHM), $\dvC$ was derived as follows.
We used FWHM of the Gaussian fit of spectra averaged over the clumps to derive the virial masses of the clumps (see below).
We derived the $\CO$($J$=1--0) line widths ($\dv_{\mathrm{obs}}$ in FWHM) by fitting the $\CO$($J$=1--0) spectra averaged over the clumps with a Gaussian function. 
We estimated intrinsic velocity widths of the clumps, $\dvC$, by correcting for the 
velocity reoslution of the instruments, $\dv_{\mathrm{spec}}$, as $\dvC$ =$\left(\dv_{\mathrm{obs}}^2-\dv_{\mathrm{spec}}^2 \right)^{1/2}$. 
We found that the velocity widths, $\dvC$ range from 1.4 to 3.3$\,$${\rm{km \: s}^{-1}}$, suggesting that non-thermal contributions are considerable ; the thermal velocity widths (one-dimensional velocity widths in FWHM) of $\CO$ molecules are 0.12--0.24$\,$${\rm{km \: s}^{-1}}$ with $T_{\rm{K}}$ = 10--30$\,$K. 
The H$_{2}$ column density of the clumps, ${N_{\rm{H}_{2}}}$ was calculated as,
\begin{eqnarray}
\label{column}
{N_{\rm{H}_{2}}} = 4.7 \times 10^{13} \left(\frac{X_{\CO}}{1.7\times 10^{-7}}\right)^{-1}
	\left(\frac{T_{\rm{ex}}}{\rm{K}}\right)
	\exp \left({\frac{5.27}{T_{\rm{ex}}/\rm{K}}}\right)
	\nonumber \\
	\quad \quad \quad
	\times	
	\left( \frac{\eta}{0.45} \right)^{-1}
	\left( \frac{\tau}{1-\exp({-\tau})} \right)
 	\left( \frac{\int {{T}^{*}_{\rm{A}}} dv}{\rm{K} \: \rm{km \: s}^{-1}} \right)
	\quad [\rm{cm}^{-2}] ,
\end{eqnarray}
where $X_{\CO}$ is the fractional abundance of $\CO$ relative to $\rm{H}_2$, 
$T_{\mathrm{ex}}$ is the excitation temperature of the transition, $\tau$ is the optical depth,
and $\int {T}^{*}_{\mathrm{A}} dv$ is the total integrated intensity of the $\CO$ line emission.
For $X_{\CO}$, we used 1.7 $\times$ $10^{-7}$ by Frerking et al. (1982).
From our $\CCO$($J$=1--0) observations with the abundance ratio of $\CO$ to $\CCO$ as 3.65 presented by Fuller $\&$ Ladd (2002), 
the optical depth of the $\CO$ emission was estimated to be 0.5$\pm$0.35, which indicates that optically thin conditions are plausible.
Given $\tau=0.5$, we confirmed ${\tau}/({1-\exp({-\tau})})$ $\sim$ 1.
We applied the result of $\tau=0.5$ for all objects. Previously, Hofner et al. (2000) revealed that most of massive star forming regions are 
optically thin condition in $\CO$ emission, so that we consider that the assumption is reasonable.
The total mass under the LTE condition, $\MCLU$ is calculated as,
\begin{eqnarray}
\label{mass}
\MCLU = 40 \left(\frac{X_{\CO}}{1.7\times 10^{-7}}\right)^{-1}
	\left( \frac{D}{2100 \: \mathrm{pc}} \right)^{2} 
	\left( \frac{\RC}{100 ''} \right)^{2} 
	\nonumber \\
	\quad \quad \quad
	\times
	\left(\frac{T_{\rm{ex}}}{\rm{K}}\right)
	\exp \left({\frac{5.27}{T_{\rm{ex}}/\rm{K}}}\right)
	\left( \frac{\eta}{0.45} \right)^{-1}
 	\left( \frac{\int {{T}^{*}_{\rm{A}}} dv}{\rm{K} \: \rm{km \: s}^{-1}} \right)
	\quad [\MO] ,
\end{eqnarray}
where $D$ is the distance to the object and $\RC$ is the radius of the clump. 
The rotation temperatures, ${T_{\rm{rot}}}$(2,2 ; 1,1) were derived from the NH$_{3}$ data using the same method presented by Ho $\&$ Townes (1983). 
We consider the kinetic temperature, $T_{\rm{K}}$ $\sim$ $T_{\rm{rot}}$ as proposed by Danby et al. (1988).
We used the kinetic temperature derived from NH$_{3}$ data as the excitation temperature of the $\CO$ gas.

\begin{figure}
\epsscale{0.8}
\plotone{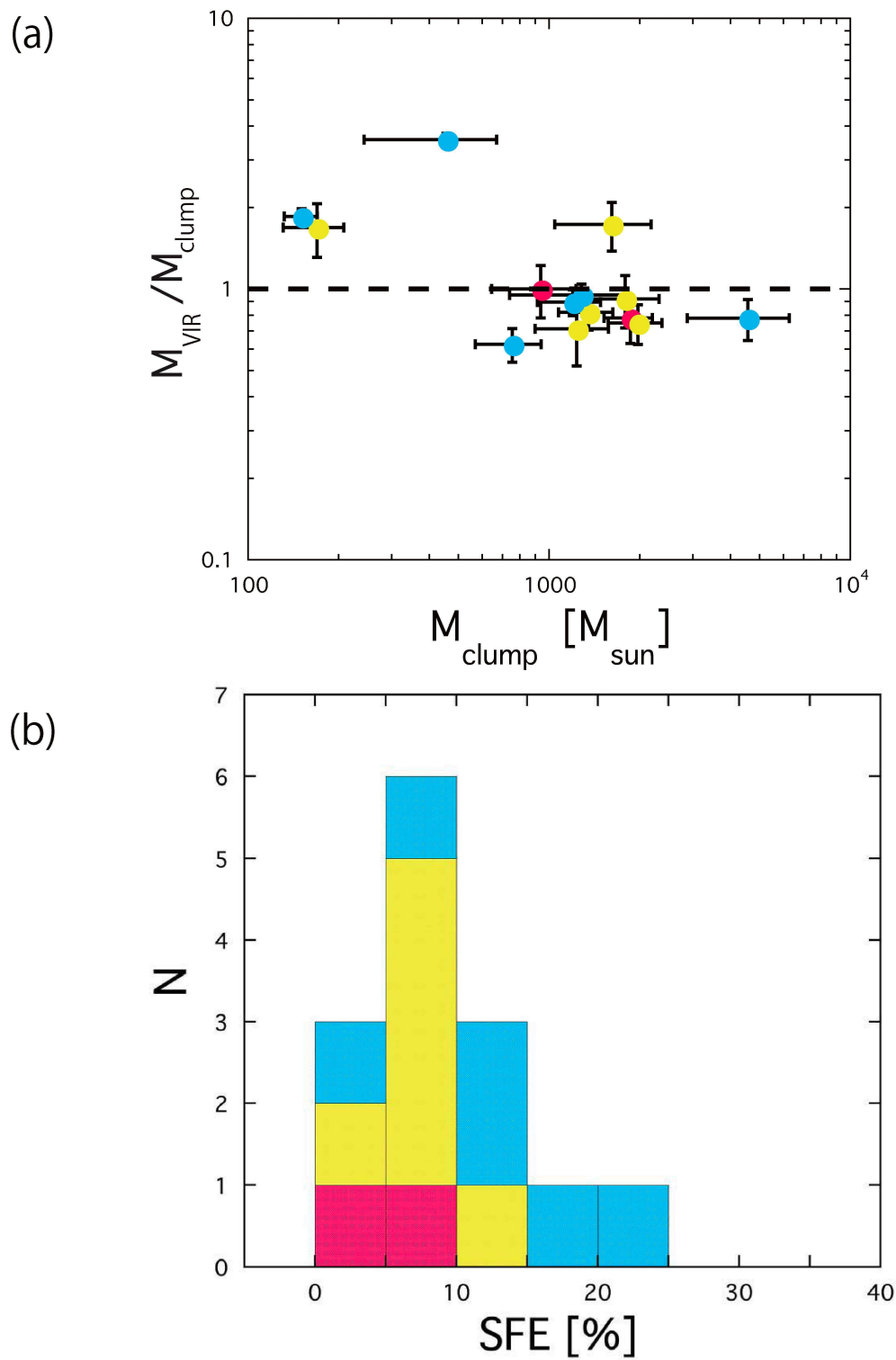}
\caption{
(a) Plot of virial ratio vs. LTE mass for our $\CO$ clumps
(pink filled circleF$\it{Type \ A}$, yellow filled circle : $\it{Type \ B}$, blue filled circle : $\it{Type \ C}$).
The dashed line indicates the virial ratio of unity.
(b) Histogram of the SFEs of $\CO$ clumps
(pink : $\it{Type \ A}$, yellow : $\it{Type \ B}$, blue : $\it{Type \ C}$).}
\label{fig2}
\end{figure}

We also calculated the virial masses of the clumps using  \\
($\MVIR/\MO)=209 (\RC/\mathrm{pc})(\dvC/\rm{km \: s}^{-1})^{2}$\citep{ike07}.
Figure $\ref{fig2}$(a) shows a plot of virial ratio vs. LTE masses of the $\CO$ clumps. 
For 70$\,$$\%$ of our targets, the virial ratios, $\MVIR/\MCLU$ are in a range of $\sim$ 1$\pm$0.5.
The ratios for AFGL 5180, Serpens, and S87E are relatively larger than the other sources.
Taking account of the accuracy of abundance ratio and temperature estimate, 
we consider that most of the $\CO$ clumps are in virial equilibrium, which indicates they are parental gas clouds for cluster formation.

\section{DISCUSSION}

\subsection{Classification of the clumps}\label{4-1}

In this section, we focus on the spatial relation between the distributions of clumps and clusters, and sort the clump-cluster systems according to them.
Our observations revealed that the $\CO$ clumps show various morphologies, compact (e.g., AFGL 5142) or extended structure 
(e.g., GGD 12-15, Gem 1), cometary (e.g., S140) or filamentary (e.g., Gem 4, BD+40$^{\circ}$4124).
Most of them have internal substructures (e.g., sub-clumps) within the clump.
Additionally, the $\CO$ maps show a variation of brightness contrast.
Under the optically thin condition of the $\CO$ line, we can consider the brightness contrast is equivalent to the column density contrast, 
which reflects the physical state of dense clumps.
There is a variation between the spatial distributions of the $\CO$ peak emissions and the cluster centers.
Extreme cases are found in two cluster samples; a stellar cluster is just located at a single peak in the distribution of the $\CO$ line emission 
(e.g.,$\,$AFGL 5142), or located at a ditch or a cavity of the emission line distribution within the clump which has multiple sub-clumps around the cavity 
(e.g.,$\,$Mon R2).
As an intermediate case between them, a cluster is associated with one of sub-clumps and, is close to the peak of the $\CO$ emission, such as GGD 12-15. 
Gem 4, in which two distinct clusters are identified, contains these two cases in agreement of locations of sub-clump peaks and clusters; one cluster is coincident at the sub-clump peak but the other is not.
The number of identified sub-clumps are dependent on the spatial resolution of observations and the distance of the target objects.
Note that for a distant object two or more sub-clumps might be identified as one component of sub-clump. 
Nevertheless, for the distance of 2.1$\,$kpc, the effective spatial resolution of our data corresponds to $\sim$ 0.22$\,$pc which is comparable to the Jeans lengths for our targets. 
Hence, the above discussion still holds with our definition and identification of the sub-clumps.

\begin{figure}
\epsscale{0.85}
\plotone{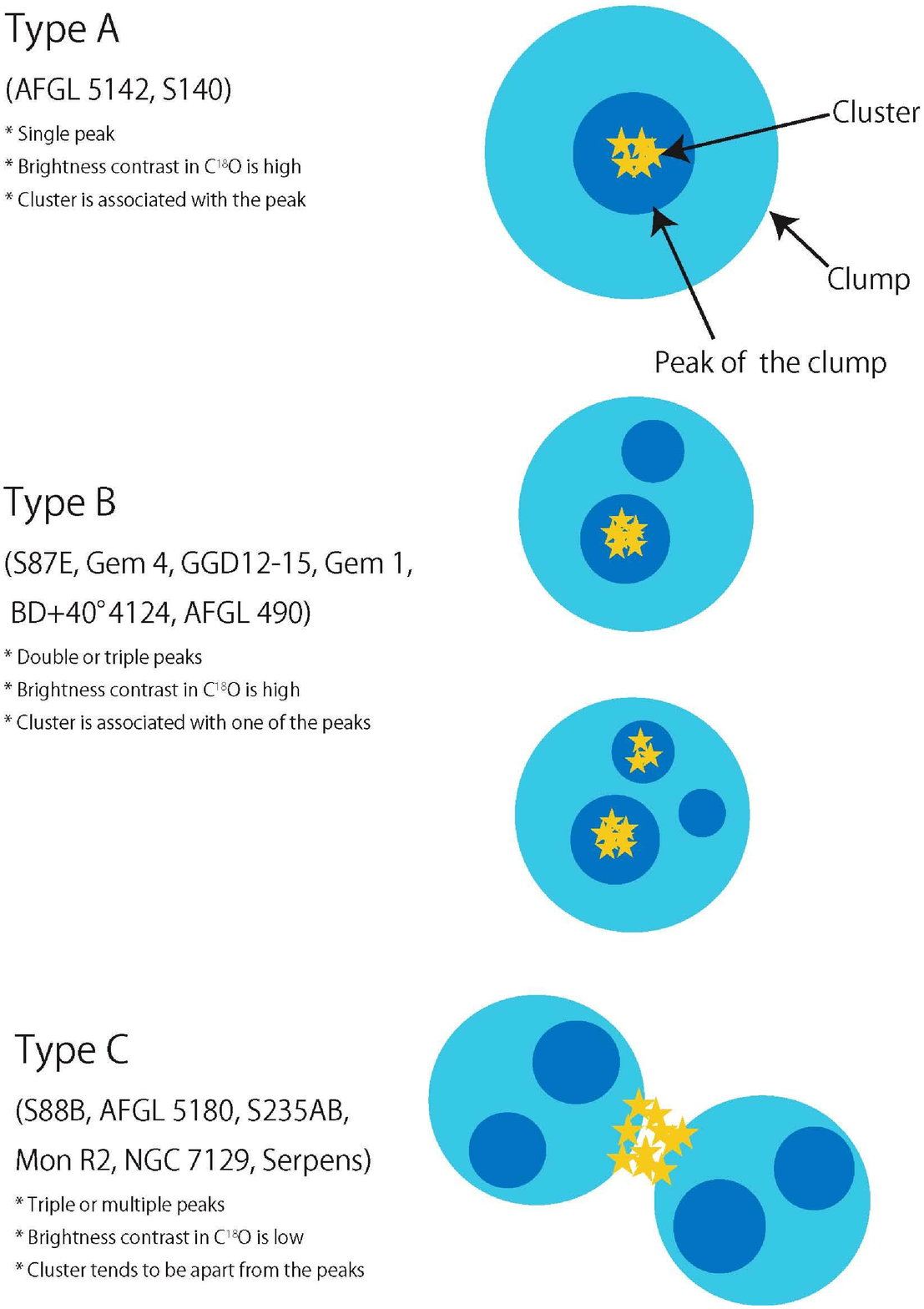}
\caption{Schematic picture of our morphological classification of cluster-forming clumps.
$\it{Type \ A}$ clumps in which the clusters are just associated with a single peak of $\CO$ emission distribution. 
These clumps have not only the single peak but higher brightness contrasts in the distributions of $\CO$ emission than others. 
$\it{Type \ B}$ clumps in which cluster is associated with one of the peaks of $\CO$ emission distribution. 
The clumps of this type have relatively higher brightness contrasts.
$\it{Type \ C}$ clumps in which the clusters are located at a cavity-like $\CO$ emission hole. 
The clumps of this type have lower brightness contrasts than those of $\it{Type \ B}$.}
\label{model}
\end{figure}

From the spatial relation between the distributions of $\CO$ emission and clusters, we sorted observed $\CO$ clumps into three types to simplify discussion and to clarify a qualitative trend.
One is the clumps in which the clusters are just associated with a single peak of $\CO$ emission distribution (we refer as $\it{Type \ A}$). 
These clumps have not only a single peak but higher brightness contrasts in the distributions of $\CO$ emission than others. 
The opposite is the clumps in which the clusters are located at a cavity-like $\CO$ emission hole, which we refer as $\it{Type \ C}$. 
The clumps of this type have lower brightness contrast than $\it{Type \ A}$.
The rest are the clumps in which a cluster is associated with one of the peaks of $\CO$ emission distribution (we refer as $\it{Type \ B}$). 
The clumps of this type have relatively higher brightness contrasts than $\it{Type \ C}$.
Considering the above features, we classified AFGL 5142 and S140 into $\it{Type \ A}$.
Here, we neglect faint components whose intensities are less than half of the peak intensity
(e.g., A faint peak in southeast area for AFGL 5142, a few components in northeast area for S140).
In contrast, we classified S88B, AFGL 5180, S235AB, Mon R2, NGC 7129, and Serpens SVS2 into $\it{Type \ C}$.
In addition, we classified S87E, Gem 4, GGD12-15, Gem 1, BD+40$^{\circ}$4124, and AFGL 490 into $\it{Type \ B}$.
For AFGL 490, the sub-clump located in bottom area is relatively faint component, so that it seems close to the condition of $\it{Type \ A}$.
Gem 4 is a rather unique one in our samples and difficult to classify into one of the three types, because this source seems to be binary clusters, i.e., a pair of bright SW and fainter NE clusters. The SW cluster is associated with a fainter one of two $\CO$ clumps, and the fainter NE cluster seems to be with another but a prominent clump. If we classify each of the clusters, SW will be $\it{Type \ A}$ and NW will be $\it{Type \ B}$ (or close to $\it{Type \ A}$). Here we classified into $\it{Type \ B}$ because the prominent $\CO$ peak and the brighter cluster are separated. However, this source will be treated as one exception of our samples for critical discussions (see $\S$ \ref{4-2-2}).
The detailed parameters of individual sources are listed in Table \ref{class} and a schematic picture
of these three types is shown in Figure \ref{model}.

\subsection{Evolution of dense gas in cluster forming regions}

\subsubsection{Progress of the cluster formation}\label{4-2-1}

In the previous section, we classified the observed $\CO$ clumps into the three types. 
Such a sequence of morphologies was recently interpreted as an indicator of evolution in cluster-forming environments \citep{fue02, rid03}. 
Especially disagreements between the locations of stellar clusters and of cloud peaks may reflect to progressive dispersal of the dense gas by embedded stars (see e.g.,$\,$Fuente et al. 2002). 
We also interpret each of our morphological classification as evolutionary stages of the cluster forming clumps (see Figure \ref{model}).

We found a tendency that along the type sequence $A$ to $C$, the clump structures become complex with some sub-clumps and local peaks.
This can be explained by fragmentations to form gas reservoirs of star formation.
In molecular clumps, natal structure of clusters, the column density structure is more or less uniform initially. Eventually fragmentation occurs with some mechanisms in the clumps and initiates star cluster formation (e.g., Bonnell et al. 2003).
The typical separation of the sub-clumps are the order of the Jeans length, which supports the interpretation.
After fragmentation, the local condensations grow to be distinct peaks where stellar clusters will be eventually born. 
During the cluster formation, the cores within the clumps form stars in a wide mass range and dense gas is dispersed by their stellar activities. Once clusters are formed, young stars, particularly most massive stars disperse surrounding gas with evolution making cavities around them (Fuente et al. 1998, 2001).
The structures of the evolved clumps are expected to be less condensed than those of the younger clumps.
For more evolved clumps, continual dispersal of dense gas around the stellar cluster makes a disagreement between the stellar positions and the peaks of gas condensations to evolve the configuration in which clusters are located at cavity-like structure of the clumps.
Fuente et al. (2002) showed that the spectral type of B0--B5 stars have enough power to disperse $\sim$ 90$\,$$\%$
of surrounding gas around the star within a radius of $\sim$ 0.1$\,$pc.
If we apply their results to our data, the dispersed masses by the cluster members which include more massive stars
are estimated to be several hundred $\MO$.
Hence, the degree of dispersal is considered to be an indicator of evolutionary stages of the cluster forming clumps, and namely corresponds to the picture of clump and cluster evolution.
We found that three objects with relatively large virial ratios (see Figure \ref{fig2} a) belong to $\it{Type \ B}$ and $\it{Type \ C}$ ($\it{Type \ C}$ : AFGL 5180 $\&$ Serpens, $\it{Type \ B}$ : S87E ), which is interpreted that dense gas is significantly dispersed due to the clump-cluster evolution.

In the above discussion, we propose that the degree of dispersal represents the evolution of clusters. We note that the degree of dispersal depends on input energy from young stars. Since massive stars are energetic, the gas dispersal are dominated by most massive stars in the cluster. 
%The most massive stars of our samples are ranging from 3 to 20$\,\MO$ ; especially 
The top seven clusters among our samples sorted with mass of the most massive stars in the cluster 
are all $\it{Type \ B}$ or $\it{C}$ and lacking $\it{Type \ A}$. 
It is emphasized that the evolution of the clumps is very sensitive to the most massive star in the clump.
%and then the discussions about the evolution of clusters in the clumps should be based on the uniform samples 
%in terms of the most massive stars and total stellar mass.
%
%
Since such a massive star has a short lifetime, the time scale of $\it{Type \ A}$ associated with very massive stars is expected to be short, which explains lack of $\it{Type \ A}$ with most massive stars in our sample. Our scenario, however, does not change because we focus on evolutionary stages (birth, evolution, and dispersion) of the physical conditions of the clump associated with the cluster, not the absolute ages of the cluster-forming clumps.

We compared our classification with that by Ridge et al. (2003) who firstly have carried out a survey of the dense clumps associated with the embedded clusters.
They suggested a scheme of the classification of the clumps and revealed that the peak positions of evolved clumps 
are apart from the Herbig Ae/Be stars or $\it{IRAS}$ sources. 
They also investigated the number of the significant intensity peaks inside the clumps and showed that evolved clumps have a tendency to have much more peaks than young clumps. 
However, their observation cannot resolve the components whose length are comparable to Jeans length.
For example, they identified Mon R2 as relatively early phase, $\it{Class \ I}$ and in fact, we cannot discern clear cavities in their $\CO$ map of Mon R2.
In our higher resolution map, however, there is a clear cavity inside the clump and there are multiple sub-clumps, which suggests that Mon R2 is relatively evolved. 
Furthermore, for BD+40$^{\circ}$4124, it is classified into $\it{Class \ I}$ in Ridge et al. (2003). 
In contrast, there are double peaks inside the clump and the peak is apart from the cluster in our results, so that we classified into $\it{Type \ B}$.
The inconsistency arises from the difference of spatial distributions of observed data.

\subsubsection{SFE along the clump-cluster evolution}\label{4-2-2}

In the previous subsection, we proposed the evolutionary scenario of dense clumps associated with young clusters.
Since a large amount of gas is dispersed by the cluster members during cluster evolution, stellar mass becomes more dominant in the cluster.
The combination of masses between the dense gas and stellar clusters is described by the Star Formation Efficiency (SFE), 
which was defined by Lada $\&$ Lada (2003).
Lada $\&$ Lada (2003) argues that the SFEs (SFE = ${M_{\rm{cluster}}}$/$(M_{\rm{cluster}}+M_{\rm{clump}})$, where $M_{\mathrm{cluster}}$ is the stellar masses of clusters and $M_{\mathrm{clump}}$ is the masses of clumps) is one of the fundamental parameters characterizing the evolution of the cluster. 
From our proposed scenario, it is expected to increase the SFEs from $\it{Type \ A}$ to $\it{Type \ C}$. 
Thus we calculated the SFEs for our targets listed in Table \ref{para}.
We found that the SFEs for the $\CO$ clumps range from 1.2 to 22$\,$$\%$ with a mean of 10$\,$$\%$,
which is as small as the values previously derived for low-mass star forming regions (e.g.,$\,$Onishi et al. 1998;$\,$Tachihara et al. 2002).
Figure $\ref{fig2}$(b) shows a histogram of the SFEs of each type and indicates that $\it{Type \ A}$ objects have smaller SFEs ($\la$ 5$\%$) than $\it{Type \ B}$ and $\it{Type \ C}$ objects ($\ga$ 10$\%$).
The mean SFEs of individual type become 3$\,$$\%$ for $\it{Type \ A}$, 7$\,$$\%$ for $\it{Type \ B}$, and 13$\,$$\%$ for $\it{Type \ C}$.
If we exclude Gem 4, the mean SFE of $\it{Type \ B}$ becomes 5$\,$$\%$, the tendency remains.
Because our classification is based on the morphologies, there is a scatter in the calculated SFEs.
For example, the SFEs for S88B and NGC 7129 are 2.5$\,$$\%$ and 9.1$\,$$\%$, respectively, 
which is smaller than the other values in spite of $\it{Type \ C}$. 
AFGL 490 is classified into $\it{Type \ B}$, but the derived SFE is as small as that of $\it{Type \ A}$, 
suggesting that AFGL 490 have a possibility to be relatively young close to $\it{Type \ A}$ as mentioned in $\S$ \ref{4-1}.
In fact, we found that $\CO$ trace not only cores but also envelopes because the previous observation revealed that the dust continuum trace 
smaller and denser areas than $\CO$ clumps (e.g., Klein et al. 2005).
Considering above uncertainties, this result implies that there is a relatively-clear tendency that SFEs increase along the evolutionary stages from $\it{Type \ A}$ to $\it{Type \ C}$ clumps.

\subsubsection{Velocity dispersion along the clump-cluster evolution}\label{4-2-3}

In this section, we focus on the relation between velocity dispersion of dense clumps and our evolutionary classification.
The second moment maps of $\CO$ show that most of our targets have relatively low and uniform velocity dispersion within the clumps as shown in Figures \ref{map1} to \ref{map6} (right) except for S87E, S88B, and Mon R2 ($\S$ \ref{3-1}). 
We found that these three objects with regions of large velocity dispersion toward the cluster centers, up to $\sim$ 1.3$\,$km s$^{-1}$ belong to $\it{Type \ B}$ and $\it{Type \ C}$ ($\it{Type \ C}$ : S88B $\&$ Mon R2,  $\it{Type \ B}$ : S87E), which is considered as the evidence to indicate that our classification roughly represents a scheme of clump-cluster evolution.
Namely, broadening the velocity dispersion is due to an integration of continuous energy input to excite the turbulence motion by stellar activities of embedded clusters. 
The three objects which have large velocity dispersions around the clusters include relatively massive cluster members in our target objects 
($\sim20\MO$ for S87E and S88B, 10$\MO$ and a few B stars for Mon R2).
Therefore, it is considered that the large velocity dispersions in the three clumps are results of energy input over a long period of evolution from active young massive sources.
Some of $\it{Type \ C}$ clumps show that the velocity dispersions are not so large as 1.3$\,$km s$^{-1}$ on this scale, 
probably because the cluster members of them are in moderate mass ranges (see Table \ref{target}).
The outflows from the cluster members are not likely to be energy sources because of the spatial disagreements between the outflow lobes and the large velocity dispersion.
The efficiency to convert the outflow energy into the turbulence energy is generally expected to be so low 
(a few $\%$ or less; Dyson $\&$ Williams (1997)) that these effects are hardly detectable unless evolved phases of cluster formation including an energetic internal source.

\section{CONCLUSION}

We have carried out a clump survey toward the 14 embedded clusters in the $\CO$ line emission with the Nobeyama 45m radio telescope.
Our aim is to reveal the evolutionary stages of the clumps, and to understand the formation and evolution of the clusters within the clumps.
Our results and conclusions are summarized as follows :

\begin{enumerate}

\item We made the $\CO$ maps with a size of $\sim$ 6$^{\prime}$ $\times$ 6$^{\prime}$ to 10$^{\prime} \times 10^{\prime}$ of the 14 nearby ($D$ $\leq$ 2.1$\,$kpc) embedded clusters.
As a result, all the clusters are found to be associated with the $\CO$ clumps, 
whose radii, masses, and velocity widths are 0.40--1.6$\,$pc, 150--4600$\,$$\MO$, and 1.4--3.3$\,$km s$^{-1}$, respectively. 
Our analysis shows that most of the clumps are likely to be in virial equilibrium.

\item We sorted the clump types based on the spatial relation between the distributions of $\CO$ emission and clusters into three types. One is the clumps in which the clusters are just associated with a single peak of $\CO$ emission distribution (we refer as $\it{Type \ A}$). These clumps have not only the single peak but higher brightness contrasts in the distributions of $\CO$ emission than others. 
The opposite is the clumps in which the clusters are located at a cavity-like $\CO$ emission hole, which we refer as $\it{Type \ C}$. 
The clumps of this type have lower brightness contrasts than those of $\it{Type \ A}$.
The rest are clumps in which cluster is associated with one of the peaks of $\CO$ emission distribution (we refer as $\it{Type \ B}$). 
The clumps of this type have relatively higher brightness contrasts than $\it{Type \ C}$.
We suggest that our morphological classification implies the evolutionary stages of cluster-forming clumps.

\item 
We found that a tendency that the SFEs of the dense clumps increase from $\it{Type \ A}$ to $\it{Type \ C}$, which also supports our scenario.
Both effects of stellar activities from massive stars and the evolutionary stages of the clumps are considered to be a cause of large velocity dispersion of the clumps (see Figure \ref{map1} to \ref{map6}). This result suggests that it takes long time to drive the turbulence from stellar activities.

\end{enumerate}

In summary, our study revealed that the results of morphological classification corresponds to the evolutionary stages of the cluster forming clumps. 
In addition, we succeed in unveiling the variations of the clumps along the evolutionary stages.
Thus, our morphological classification can well extend the classification for a single star by Fuente et al. (1998a, 2002) to cluster forming clumps. 
It is suitable for investigating the evolutionary stages of the clumps and reasonable perspective to reveal the initial conditions of cluster formation, 
as a step in understanding the formation processes of the clusters.

\acknowledgments

We thank the referee for the constructive comments that have helped to improve this manuscript.
We also thank Drs.$\,$Yoshimi Kitamura and Kazuyoshi Sunada for useful discussion about our data.
We are grateful to the staff of the Nobeyama Radio Observatory (NRO)$\footnote{Nobeyama Radio Observatory is a branch of the National Astronomical Observatory of Japan, National Institutes of Natural Sciences.
}$ for both operating the Nobeyama 45m telescope and helping us with the data reduction.
In particular, we acknowledge Drs.$\,$Nario Kuno, Shuro Takano, Tsuyoshi Sawada, Takeshi Sakai, and Norio Ikeda for their contributions to our observations.

\appendix

\section{Individual sources}\label{sec2}

\subsection{S87E}

S87E is located at a distance of 2.1$\,$kpc \citep{cle85}. 
There is a CO($J$=1--0) cloud with a diameter of 2$'$ and mass of 500-1000$\,$$\MO$ \citep{bl83}.
There is a compact H{\sc ii} region whose size is $\sim$ 0.6$\,$pc \citep{sha59}.
Chen et al. (2004) estimated the most massive stellar mass is $\sim$ 20$\,$$\MO$.
There is a massive bipolar outflow \citep{bar89}, which is oriented northeast-southwest region.
We found that the $\CO$ clump covers the whole cluster without apparent cavities.
There are three sub-clumps in northeast area (Hereafter, we name the sub-clumps using the directions, like S87E-NE), 
east one (S87E-E), and southwest one (S87E-SW) within the clump.
The central region of the cluster is associated with S87E-SW, while the other sub-clumps are not associated with a cluster.
The velocity dispersion of the central area of the clump is $\sim$1.1$\,$km {s}$^{-1}$, which is larger than other sources.

\subsection{S88B}

S88B is located at a distance of 2$\,$kpc.
There is a CO($J$=1--0) cloud whose mass is $\sim$ 5200$\,$$\MO$ \citep{goe99}.
There exist two compact H{\sc ii} regions (Deharveng et al. 2000 ; Goetz et al. 2003).
The most massive stellar mass is $\sim$ 20$\,$$\MO$ \citep{gar93}. 
There is a CO bipolar outflow which was detected by Phillips $\&$ Mampaso (1991).
We found that the $\CO$ clump is associated with the cluster but there is a small cavity around the cluster whose size is $\sim$0.1$\,$pc.
The prominent sub-clump (S88B-E) is located 1$'$ east of the central region of the cluster.
In addition, there are the other two sub-clumps (S88B-N, S88B-NW).
The velocity dispersion of the central area of the clump is $\sim$1.3$\,$km {s}$^{-1}$, which is larger than other clusters.

\subsection{Gem 4 (AFGL 6366S)}

Gem 4 and AFGL 6366S are located at a distance of 1.5$\,$kpc in the Gem OB1 cloud.
There is a CS($J$=2--1) clump whose mass is $\sim$ 1400 $\MO$ \citep{car95b}.
The most massive stellar mass in the cluster is $\sim$18$\,$$\MO$ \citep[e.g.,][]{gho00}.
There is a CO bipolar outflow which has been detected by Snell et al. (1988).
We found that the $\CO$ clump covers the whole clusters without apparent cavities.
The clump has filamentary structure and it is extended along the southwest-northeast direction.
There are two sub-clumps in southwest area (Gem 4-SW) and northeast one (Gem 4-NE).
The Gem 4-SW is associated with clusters, while the Gem 4-NE is shifted by 0$'$.5 from the cluster.
The velocity dispersion of the central area of the clump is $\sim$0.7$\,$km {s}$^{-1}$.
There are no correlations between the areas of large velocity dispersion and the outflow lobes.

\subsection{AFGL 5180}

AFGL 5180 is located at a distance of 1.5$\,$kpc in the Gem OB1 cloud and east area within 10$'$ of Gem 4 (AFGL 6366S).
There is a CS($J$=2--1) clump whose mass is $\sim$ 990$\,$$\MO$ \citep{car95b}.
The most massive stellar mass in the cluster is $\sim$18$\,$$\MO$ \citep[e.g.,][]{min05}.
Snell et al. (1988) surveyed outflows in CO line emission, but they did not detect any clear outflows.
We found that the $\CO$ clump covers the whole cluster without apparent cavities.
The clump is extended along the southwest-northeast direction.
There are at least two sub-clumps in the southwest area (AFGL 5180-SW) and the northeast one (AFGL 5180-NE).
The cluster is located between AFGL 5180-SW and AFGL 5180-NE.
The velocity dispersion of the central area of the clump is $\sim$0.5$\,$km {s}$^{-1}$.

\subsection{GGD12-15}

GGD 12-15 is located at a distance of 830 pc in the Monoceros molecular cloud.
There is a $^{13}\rm{CO}$($J$=1--0) clump whose mass is $\sim$ 1500$\,$$\MO$ \citep{rid03}.
There is a cometary compact H{\sc ii} region \citep{gom98} at the central position of the cluster.
The most massive stellar mass in the cluster is $\sim$15$\,$$\MO$ \citep[e.g.,][]{fan04}.
There is a CO bipolar outflow which has been detected by Little et al. (1990).
We found that the $\CO$ clump covers the whole cluster without apparent cavities.
The clump has spherical shape.
There are at least four sub-clumps in the central region of the clump (GGD12-15-Main), and
there is another sub-clump in north-west area (GGD12-15-NW).
The velocity dispersion of the central area of the clump is $\sim$0.6$\,$km {s}$^{-1}$.
There are no correlations between the areas of large velocity dispersion and the outflow lobes.

\subsection{S235AB}

S235AB is located in the Perseus spiral arm, at a distance of 1.8$\,$kpc \citep{sha59}.
There is a $^{13}\rm{CO}$($J$=1--0) clump whose mass is $\sim$4200$\,$$\MO$ \citep{nak86}.
The most massive stellar mass in the cluster is $\sim$15$\,$$\MO$ \citep[e.g.,][]{fel97}.
There is an optical H{\sc ii} region which is ionized by the most massive star.
Nakano $\&$ Yoshida (1986) detected a CO molecular outflow which is aligned in a north-east to south-west direction.
We found that the $\CO$ clump covers the whole cluster without apparent cavities.
The clump is extended along the southeast-northwest direction.
There are mainly two sub-clumps in the central region of the clump (S235AB-N, S235AB-S).
S235AB-N is associated with the cluster, while S235AB-S is not.
The velocity dispersion of the central area of the clump is smaller than $\sim$0.6$\,$km {s}$^{-1}$.
There are no correlations between the areas of large velocity dispersion and the outflow lobes.

\subsection{Gem 1}

Gem 1 is located at a distance of 1.5 kpc in the Gem OB1 cloud and north-east area of Gem4 and AFGL 6366S.
There is a CS($J$=2--1) clump whose mass is $\sim$ 310$\,$$\MO$ \citep{car95b}.
The most massive stellar mass in the cluster is $\sim$13$\,$$\MO$ \\ \citep[e.g.,][]{ana04}.
Snell et al. (1988) observed CO line emission towards Gem1 but they have not detected any outflows.
We found that the $\CO$ clump covers the whole cluster without apparent cavities.
The clump is elliptical structure.
There are two sub-clumps in northwest area (Gem 1-NW) and southeast one (Gem 1-SE).
Gem 1-NW is associated with the cluster, while Gem 1-SE is not.
The velocity dispersion of the central area of the clump is $\sim$0.8$\,$km {s}$^{-1}$.
There is an areas with large velocity dispersion 1$'$ apart from central region of the cluster.

\subsection{AFGL 5142}

AFGL 5142 is located at a distance of 1.8$\,$kpc \citep{sne88}. 
There is a NH$_{3}$ clump whose mass is $\sim$700$\,$$\MO$ \citep{zha02}.
There is a ultra compact H{\sc ii} region in the central region of the cluster.
The most massive stellar mass in the cluster is $\sim$10$\,$$\MO$ \citep{zha02}.
A CO outflow associated with the star was found by Hunter et al. (1995).
We found that the $\CO$ clump covers the whole cluster without apparent cavities.
The clump is upside-down triangle structure.
The peak of the clump is associated with the central region of the cluster.
The degree of central concentration of this clump is higher than other sources.
The velocity dispersion of the central area of the clump is $\sim$0.5$\,$km {s}$^{-1}$.
There are no correlations between the areas of large velocity dispersion and the outflow lobes.

\subsection{Mon R2}

Mon R2 is located at a distance of 830 pc in the Monoceros molecular cloud.
There is a $^{13}\rm{CO}$($J$=1--0) clump whose mass is $\sim$2600$\,$$\MO$\citep{rid03}.
At the center of the clump, there is a compact H{\sc ii} region.
There is a powerful molecular outflow \citep{wol90, taf97}.
The most massive star in the cluster is 10$\,$$\MO$ \citep{car97}.
We found that the $\CO$ clump covers the whole cluster, in addition there are multiple sub-clumps inside the clump.
The clump is elliptical structure.
The molecular gas is absent in the east and west area of the map.
The brightness contrast of the clump is lower than other sources. 
The velocity dispersion of the central area of the clump is $\sim$1.3$\,$km {s}$^{-1}$, which is larger than other objects.
The well studied large outflow is located to the center of the clump, the blue-shift sub-clump is consistent with the areas of large velocity dispersion.

\subsection{S140}

S140 is located in the molecular cloud L1204 at a distance of 910 pc (Crampton $\&$ Fisher 1974).
There is a $^{13}\rm{CO}$($J$=1--0) clump whose mass is $\sim$ 1600$\,$$\MO$ \citep{rid03}.
IRS 1 in S140 is the brightest member of the clusters, whose mass is 10$\,$$\MO$ \citep{pre02}.
There is a CO bipolar outflow extended along the northwest-southeast direction \\ \citep{hay87}.
We found that the $\CO$ clump covers the whole cluster without apparent cavities.
The clump is cometary structure.
There is a single peak within the clumps.
The brightness contrast of the clump is higher than other sources.
The velocity dispersion of the central area of the clump is $\sim$0.9$\,$km {s}$^{-1}$.
There are no correlations between the areas of large velocity dispersion and the outflow lobes.

\subsection{AFGL 490}

AFGL 490 is located in the Cam OB1 complex \citep{chi91} at a distance of 900$\,$pc.
There is a $^{13}\rm{CO}$($J$=1--0) clump whose mass is $\sim$ 920$\,$$\MO$ \citep{rid03}.
It has a powerful CO bipolar outflow which is oriented roughly northeast-southwest \citep{mit95}.
There is no observational evidence of an H{\sc ii} region.
The most massive star is Herbig Be stars whose mass is $\sim$8$\MO$ \citep{sch02}.
We found that the $\CO$ clump covers the whole cluster without apparent cavities.
The clump is filamentary structure.
There are two sub-clumps within the clumps (AFGL 490-N, AFGL 490-S).
Only AFGL 490-N is associated with the cluster.
The velocity dispersion of the central area of the clump is $\sim$0.5$\,$km {s}$^{-1}$.
There are no correlations between the areas of large velocity dispersion and the outflow lobes.

\subsection{BD$+$40$^{\circ}$4124}

BD$+$40$^{\circ}$4124 is located in the Cygnus arm at a distance of 1$\,$kpc.
There is a $^{13}\rm{CO}$($J$=1--0) clump whose mass is $\sim$ 230$\,$$\MO$ \citep{rid03}.
BD$+$40$^{\circ}$4124 is the brightest member and whose mass is 8$\,$$\MO$ \citep{loo06}.
The molecular outflow has been detected in CO($J$=2--1) molecular line which is located at the position of the southern source, V1318 Cyg \citep{pal95}.
There is no observational evidence for a H{\sc ii} region \citep{ski93}.
We found that the $\CO$ clump covers the whole cluster without apparent cavities.
The clump is filamentary structure.
There are two sub-clumps within the clumps (BD$+$40$^{\circ}$4124-N, BD$+$40$^{\circ}$4124-S).
Only BD$+$40$^{\circ}$4124-S is associated with the cluster.
The velocity dispersion of the central area of the clump is less than $\sim$0.3$\,$km {s}$^{-1}$, which is smaller than other objects.
There are no correlations between the areas of large velocity dispersion and the outflow lobes.

\subsection{NGC 7129}

NGC 7129 is located at a distance of 1$\,$kpc \citep{rac68}.
There is a $^{13}\rm{CO}$($J$=1--0) clump whose mass is $\sim$ 1000$\,$$\MO$ \citep{rid03}.
There is no observational evidence of a H{\sc ii} region.
There are two intermediate stars, whose masses are 8$\MO$ and 6$\,$$\MO$, respectively \citep{fue02}.
Many other Herbig-Haro objects have been detected around the region that are associated with outflows \citep{eir98}.
The clump is filamentary structure.
There are mainly two sub-clumps in north area (NGC 7129-N) and south one (NGC 7129-S).
In addition, there are multiple structures inside the sub-clumps.
NGC 7129-N is associated with the cluster, while NGC 7129-S is not.
The brightness contrast of the clump is lower than other sources.
The velocity dispersion of the central area of the clump is less than $\sim$0.7$\,$km {s}$^{-1}$, which is smaller than other objects.
There are no correlations between the areas of large velocity dispersion and the outflow lobes.

\subsection{Serpens}

Serpens SVS2 is located in the Serpens molecular cloud at 260--310$\,$pc. There is a CO cloud whose mass is $\sim$ 1500$\,$$\MO$ \citep{whi95}.
The protocluster radius ($\sim$0.2$\,$pc) and the mean stellar density (400-800$\,$pc stars {pc}$^{-3}$).
There is no observational evidence of an H{\sc ii} region.
The most massive star within the cluster is SVS2, whose mass is $\sim$3$\,$$\MO$ \citep{kaa04}.
There are numerous molecular outflows \citep{dav99}.
We found that the $\CO$ clumps cover the whole clusters without apparent cavities.
The clump is filamentary structure.
There are two sub-clumps within the clumps (Serpens-S, Serpens-NW).
In addition, the Serpens-S have two sub-clumps inside area.
The cluster is located between Serpens-S and Serpens-NW.
The velocity dispersion of the central area of the clump is $\sim$0.5$\,$km {s}$^{-1}$.
There are no correlations between the areas of large velocity dispersions and the outflow lobes.

\clearpage

\clearpage

\clearpage

\clearpage

\clearpage

\clearpage

\clearpage

\clearpage

\begin{deluxetable}{l l l l l l l c c c c c c c c}
\tabletypesize{\scriptsize}
\rotate
\tablecaption{Properties of our target embedded clusters\label{target}}
\tablewidth{0pt}
\tablehead{\colhead{Source Name} & \colhead{RA (J2000)} & \colhead{Dec (J2000)} & 
{$D$ [pc]} & \colhead{$N_{\rm{star}}$$\tablenotemark{a}$} & \colhead{$M_{\rm{cluster}}$ [$\MO$]} & 
\colhead{H{\sc ii} region$\tablenotemark{b}$} & \colhead{$M_{\rm{star-max}}$$\tablenotemark{c}$ [$\MO$]} & \colhead{References} }
\startdata
 S87E & 19:46:19.9 & 24:35:24 & 2100 & 101 & 180 & CH{\sc ii},UCH{\sc ii} & 20$\pm$4 & 1 \\
 S88B &  19:46:47.0 & 25:12:43 & 2000 & 98 & 120 & CH{\sc ii} & 20$\pm$4 & 2  \\
 Gem 4 &  06:08:41.0 & 21:30:49 & 1500 & 114 & 190 & H{\sc ii}& 19$\pm$4 & 3 \\
 %(AFGL 6366S & 06:08:40.9 & 21:31:00 & 1500 & 550 & 300 & -- & -- & 3) $\tablenotemark{d}$ \\
 AFGL 5180 & 06:08:54.1 & 21:38:24 & 1500 & 94 & 60 & H{\sc ii}& 18$\pm$4 & 4 \\
 GGD 12-15 & 06:10:50.9 & $-$06:11:54 & 830 & 134 & 73 & CH{\sc ii}& 15$\pm$3 & 5\\
 S235AB  &  05:40:52.5 & 35:41:25 & 1800 & 300 & 220 & H{\sc ii}& 15$\pm$3 & 6 \\
 Gem 1 &  06:09:05.4 & 21:50:20 & 1500 & 56 & 95 & UH{\sc ii}& 13$\pm$3 & 7 \\
 AFGL 5142 &  05:30:45.6 & 33:47:51& 1800 & 60 & 50 & UCH{\sc ii} & 10$\pm$2 & 8 \\
 Mon R2 & 06:07:46.6 & $-$06:22:59 & 830 & 371 & 340 & CH{\sc ii} & 10$\pm$2 & 9\\
 S140 & 22:19:18 & 63:18:48 & 900 & 16 & 12 & -- & 10$\pm$2 & 10 \\
 AFGL 490 &  03:27:38.7 & 58:46:58 & 910 & 45 & 25  & -- & 8$\pm$2 & 11 \\
 BD$+$40$^{\circ}$4124  & 18:29:56.8 & 01:14:46 & 900 & 33 & 12  & -- & 8$\pm$2 & 12  \\
 NGC 7129 & 21:43:02 & 66:06:29 & 1000 & 139 & 76 & -- & 8$\pm$2 & 13  \\
 Serpens SVS2 & 18:29:56.8 & 01:14:46 & 260 & 51 & 27 & -- & 3$\pm$1 & 14 \\ 
\hline

\enddata 
\tablerefs{
(1) Chen et al. (2004) ;
(2) Garay et al. (1993) ;
(3) Ghosh et al. (2000) ;
(4) Minier et al. (2005) ;
(5) Fang $\&$ Yao (2004) ;
(6) Felli et al. (1997) ;
(7) Anandarao et al. (2004) ;
(8) Zhang et al. (2002) ;
(9) Carpenter (2000) ;
(10) Preibisch \& Smith(2002) ;
(11) Schreyer et al. (2002) ;
(12) Looney et al.(2006) ;
(13) Fuente et al. (2002) ;
(14) Kaas et al. (2004)}
{\tablecomments{This table shows the name, center coordinates, distance, number of the member stars, mass (from Lada $\&$ Lada (2003), Porras et al. (2003)), size of the associated H{\sc ii} regions, and maximum mass of the member stars for our target clusters.}}

\tablenotetext{a}{The total stellar number of the cluster}
\tablenotetext{b}{The size of associated H{\sc ii} regions. CH{\sc ii}: Compact H{\sc ii} region, UCH{\sc ii}: Ultra compact H{\sc ii} region}
\tablenotetext{c}{The highest stellar mass of the cluster members}
%\tablenotetext{d}{Gem4 and AFGL 6366S are located within the same $\CO$ clumps}

\end{deluxetable}

\clearpage

\begin{deluxetable}{l l l l l l c c c c c c c}
\tabletypesize{\scriptsize}
\rotate
\tablecaption{Physical parameters of the $\CO$ clumps
\label{para}}
\tablewidth{0pt}
\tablehead{
\colhead{Source Name} & \colhead{$\RC$[pc]} & \colhead{$\dvC$[km $\mathrm{s}^{-1}$]} & 
\colhead{$\MVIR$ [$\MO]$} & \colhead{$N(\mathrm{H}_{2}$)[$\times$ 10$^{22}$ $\mathrm{cm}^{-2}$]} & \colhead{$\MCLU$[$\MO$]} & 
\colhead{$T_{\rm{ex}}$ [$K$]}}
\startdata
 S87E & 1.2$\pm$0.6 & 3.3$\pm$0.13 & 2800$\pm$1600 & 1.7$\pm$0.2 & 1600$\pm$570 & 22  \\
 S88B & 1.6$\pm$0.8 & 3.3$\pm$0.13 & 3600$\pm$1900 & 3.0$\pm$0.4 &  4600$\pm$1700 & 32  \\
 Gem 4 & 1.2$\pm$0.5 & 1.9$\pm$0.15  & 880$\pm$480 & 1.5$\pm$0.1 & 1200$\pm$340 & 28  \\
 AFGL 5180 & 1.3$\pm$0.6 & 2.4$\pm$0.15 & 1600$\pm$850 & 0.45$\pm$0.1 & 460$\pm$220 & 24  \\
 GGD 12-15 & 1.1$\pm$0.3 & 2.2$\pm$0.10 & 1100$\pm$380 & 1.9$\pm$0.2 & 1400$\pm$280 & 19  \\
 S235AB  & 1.3$\pm$0.6 & 2.1$\pm$0.11  & 1200$\pm$640 & 1.3$\pm$0.3 & 1300$\pm$550 & 24  \\
 Gem 1 & 1.2$\pm$0.5 & 2.6$\pm$0.15  & 1600$\pm$850 & 2.1$\pm$0.2 & 1800$\pm$540 & 26  \\
 AFGL 5142 & 1.2$\pm$0.6 & 1.9$\pm$0.12 & 940$\pm$500 & 1.0$\pm$0.1 & 900$\pm$300 & 17  \\
 Mon R2 & 1.1$\pm$0.3 & 2.2$\pm$0.10 & 1100$\pm$550 & 1.8$\pm$0.2 & 1200$\pm$290 & 19  \\
 S140 & 1.0$\pm$0.3 & 2.6$\pm$0.12 & 1500$\pm$520 & 2.9$\pm$0.2 & 1900$\pm$340 & 24  \\
 AFGL 490 & 1.4$\pm$0.5 & 2.3$\pm$0.11 & 1500$\pm$540 & 1.7$\pm$0.2 & 2000$\pm$400 & 16  \\
 BD+40$^{\circ}$4124 & 0.70$\pm$0.12 & 1.4$\pm$0.10 & 290$\pm$130 & 0.7$\pm$0.1 & 210$\pm$50 & 15$\tablenotemark{a}$  \\
 NGC 7129 & 0.90$\pm$0.3 & 1.6$\pm$0.11 & 470$\pm$180 & 1.4$\pm$0.2 & 760$\pm$190 & 12  \\
 Serpens SVS2 & 0.40$\pm$0.1 & 1.9$\pm$0.10 & 280$\pm$55 & 1.8$\pm$0.2 & 150$\pm$20 & 14  \\

\hline

\enddata
{\small
\tablecomments{
$\RC$ : radius, $\dvC$ : velocity width, $\MVIR$ : virial mass,
$N(\mathrm{H}_{2})$ : mean $\mathrm{H}_{2}$ column density, $\MCLU$: clump mass,  $T_{\rm{ex}}$ : excitation temperature derived by kinetic temperature of NH$_{3}$}}
\tablenotetext{a}{Because of the low signal data of the observation, we adopted 15$\,\rm K$ which is the typical temperature
in intermediate regions.}
\end{deluxetable}

\clearpage

\begin{deluxetable}{l l l l l l l l c c c c c c c c c}
\tabletypesize{\scriptsize}
\rotate
\tablecaption{Score sheet of the morphological classification of $\CO$ clumps
\label{class}}
\tablewidth{0pt}
\tablehead{
\multicolumn{0}{c}{Source Name} & \colhead{Type} & \colhead{SFE} & \colhead{Virial ratio} & \colhead{The number of} & \colhead{Brightness contrast} & \colhead{Offset between}  & \colhead{The largest velocity dispersion} \\
\colhead{} & \colhead{} & \colhead{[$\%$]} & \colhead{[$\MVIR/\MCLU$]} & \colhead{sub-clumps}  & \colhead{}  & \colhead{clumps and clusters [pc]}
& \colhead{of 2nd moment map [km $\rm{s}^{-1}$]} }

\startdata
 S87E@& B & 10$\pm$3 & 1.7$\pm$0.4 & 3 & High & Peak $\&$ Off-peak : 0.6--1 & 1.1 \\
 S88B & C & 2.5$\pm$0.9 & 0.8$\pm$0.1 & 4 & Low & Off-peak : 0.6--1.8 $\&$ Cavity : $\sim$ 0.5 & 1.3 \\
 Gem 4 & B & 13$\pm$3 & 0.7$\pm$0.2 & 2 & High & Peak $\&$ Off-peak : $\sim$ 0.5  & 0.7 \\
 AFGL 5180 & C & 12$\pm$5 & 3.5$\pm$0.2 & $>$ 3 & Low & Off-peak : 0.2--0.6  & 0.5 \\
 GGD 12-15 & B & 5.1$\pm$1 & 0.8$\pm$0.1 & $\sim$4 & High & Peak $\&$ Off-peak : 0.3-0.5  & 0.6 \\
 S235AB & C & 15$\pm$5 & 0.9$\pm$0.1 & $>$ 3 & Low & Off-peak : 0.3--0.5 & 0.6  \\
 Gem 1 & B & 5.0$\pm$1 & 0.9$\pm$0.2 & 2 & Low & Peak $\&$ Off-peak : $\sim$ 1 & 0.8 \\
 AFGL 5142 & A & 5.0$\pm$2 & 1.0$\pm$0.2 & 1 & High & Peak & 0.5 \\
 Mon R2 & C & 22$\pm$4 & 0.9$\pm$0.1 & $>$ 5 & Low & Off-peak : 0.3--0.7 $\&$ Cavity : $\sim$ 0.3  & 1.3\\
 S140 & A & 1.2$\pm$0.2 & 0.8$\pm$0.2 & 1 & High  & Peak  & 0.9 \\
 AFGL 490 & B & 1.2$\pm$0.3 & 0.7$\pm$0.1 & 2 & High & Peak $\&$ Off-peak : $\sim$ 1  & 0.5 \\
 BD+40$^{\circ}$4124  & B & 5.3$\pm$1 & 1.4$\pm$0.3 & 2 & High & Peak $\&$ Off-peak : $\sim$ 0.5  & 0.3 \\
 NGC 7129 & C & 9.1$\pm$2 & 0.6$\pm$0.1 & $>$ 6 & Low & Off-peak : 0.3--1.2 $\&$ Cavity : $\sim$ 0.5 & 0.7 \\
 Serpens SVS2 & C & 15$\pm$2 & 1.8$\pm$0.1 & $>$ 4 & Low & Off-peak : 0.2--0.5  & 0.5 \\

\hline
\enddata
{\small
\tablecomments{
This table shows the source name, the classified type of the clump, the SFE derived by using the $\CO$ mass, virial ratio ($\MVIR/\MCLU$) of the clump, the number of sub-clumps, the brightness contrast of the clump, the location of the cluster, and the largest velocity dispersion of 2nd moment map.
The location of the cluster is written as follows,
Peak : The emission peak of the clump which is associated with the cluster, the spatial separation between the $\CO$ peak position and the cluster center is smaller than the Jean length ;
Off-Peak : The $\CO$ emission peak of the clump which is not associated with the cluster. The separation scale between the $\CO$ peak position and the cluster center is subsequently-shown [pc] ; 
Cavity : In case that the cluster is located in a cavity. The size of the cavity is subsequently-shown [pc]. }}
\end{deluxetable}

\clearpage

\clearpage

\end{document}